\documentclass[aps,preprint]{revtex4}
\def\beq{\begin{equation}} \def\eeq{\end{equation}}
\usepackage{graphics,epsfig,amssymb}

\begin{document}

\title{Isovector proton-neutron pairing and Wigner energy in Hartree-Fock mean field calculations}

\author{D. Negrea and N. Sandulescu 
\footnote{corresponding author, email: sandulescu@theory.nipne.ro}}
\affiliation{
National Institute of Physics and Nuclear Engineering, P.O. Box MG-6, 76900 Bucharest-Magurele, Romania}

\begin{abstract}
We propose a new approach for the treatment of isovector pairing in
self-consistent mean field calculations  which conserves exactly the 
isospin and the particle number in the pairing channel. The mean field 
is generated by a Skyrme-HF functional while the isovector pairing 
correlations are described in terms of quartets formed by two neutrons 
and two protons coupled to the total isospin T=0. In this framework we analyse 
the contribution of isovector pairing to the symmetry and Wigner energies. 
It is shown that the isovector pairing provides a good description of
the Wigner energy, which is not the case for the mean field calculations 
in which the isovector pairing is treated by  BCS-like models.

\end{abstract}

\maketitle


In the last years a lot of effort has been dedicated to the understanding of 
the role played 
by the  proton-neutron pairing in the binding energies of  nuclei with $N \approx Z$. 
The experimental masses indicate  that the nuclei with $N = Z$  have an additional
binding compared to the  neighbouring nuclei. In the phenomenological mass formulas
\cite{moller} this additional binding energy is taken into account through
a term proportional to $|N-Z|$, called usually the Wigner energy.
In the  extensive mean-field calculations of 
nuclear masses the Wigner energy cannot be accounted for and therefore it is  
added as an {\it ad hoc}  phenomenological term \cite{goriely}.  

There is a long debate about the origin of the Wigner 
energy (e.g., see \cite{neergard} and references quoted therein). 
In some  studies it is supposed that the Wigner energy originates from the
proton-neutron pairing correlations, which become stronger in N=Z nuclei.
Thus, it was recently argued that the isovector proton-neutron pairing can
describe most of the extra binding associated to the Wigner energy,
provided the isovector pairing is treated beyond the BCS approximation  
\cite{bentley1,bentley2}.

 The fact that BCS-like models are not appropiate for calculating the contribution
of the isovector pairing correlations to Wigner energy can be clearly seen when
they are applied for a degenerate shell. In this case it can be analitically 
shown (e.g., see Ref.\cite{dobes}) that the BCS  approximation for the 
isovector pairing does not predict for the binding energy a linear term 
in $T_z=|N-Z|/2$, specific to the Wigner energy. On the other hand, in the 
exact solution
a linear term in isospin appears naturally through the dependence 
of energy on T(T+1),
 which reflects the isospin invariance of the isovector pairing interaction.

There is also a more general argument which indicates that the   
BCS-like models do  not  
describe properly the isovector proton-neutron pairing correlations in nuclei.
Thus, the BCS equations for the isovector pairing in N=Z nuclei have two 
degenerate solutions, one corresponding to $\Delta_n = \Delta_p \neq 0 $ 
and $\Delta_{np} = 0$ 
and the other to $\Delta_n = \Delta_p = 0 $ and $\Delta_{np} \neq 0$
 \cite{ginocchio,sandulescu_errea}.  This means that
in the BCS approximation the isovector proton-neutron pairing  
does not coexist with the like-particle pairing, as one would expect 
from the isospin symmetry. Moreover, as shown in Ref.\cite{ginocchio}, 
BCS predicts no isovector proton-neutron pairing correlations
in the ground state of $N>Z$ nuclei with $T=|T_z|$. One reason
why BCS fails to describe properly the isovector proton-neutron correlations 
is because 
it does not  conserve exactly the particle  number and the isospin. 
The two symmetries can be exactly restored performing projected-BCS (PBCS) calculations.  
However, the comparison with exactly solvable models shows
that  PBCS is still unable to provide acurate results for the isovector
pairing correlations \cite{qcm1,chen},
which demonstrates the need of going beyond the BCS-type models.  
In Refs.\cite{qcm1,qcm2} it was proved that an approach based on quartets formed by two
neutrons and two protons coupled to the total isospin T=0 can describe with very high
accuracy the isovector pairing correlations in the ground state of both N=Z and 
$N>Z$ nuclei. 
In this paper we  show how this approach can be applied for treating 
accurately the isovector
pairing  in self-consistent Hartree-Fock (HF) mean field models.  Then, within 
this framework, we  analyse
the contribution of the isovector proton-neutron pairing to the symmetry and Wigner
energies. 
 
For consistency reason we start by presenting briefly the quartet model
introduced in Refs.\cite{qcm1,qcm2}. This model describes the ground state of  a 
system formed by N neutrons and Z protons moving outside a self-conjugate core and 
interacting via  an isovector pairing force. The Hamiltonian describing this system is 
\beq
\hat{H}= \sum_{i,\tau=\pm 1/2} \varepsilon_{i\tau} N_{i\tau}
-g \sum_{i,j,t=-1,0,1}  P^+_{i,t} P_{j,t} ,
\eeq
where $\varepsilon_{i\tau}$ are the single-particle energies associated
to the mean fields of neutrons ($\tau=1/2$) and protons ($\tau=-1/2$), 
supposed invariant to time reversal. The isovector interaction is expressed 
in terms of the isovector pair operators 
$P^+_{i,1}=\nu^+_i \nu^+_{\bar{i}}$, $P^+_{i,-1}=\pi^+_i \pi^+_{\bar{i}}$ and 
$P^+_{i,0}=(\nu^+_i \pi^+_{\bar{i}} + \pi^+_i \nu^+_{\bar{i}})/\sqrt{2}$; the
operators $\nu^+_i$ and $\pi^+_i$ create, respectively, a neutron and a proton in 
the state $i$ while $\bar{i}$ denotes the time conjugate of the state $i$.

Following Ref.\cite{qcm1}, the ground state of  Hamiltonian (1) for a system with N=Z
is described by the state
\begin{equation}
| \Psi \rangle =(A^+)^{n_q} |0 \rangle ,
\end{equation}
where $n_q=(N+Z)/4$ and $A^+$ is the collective quartet built by two isovector pairs
coupled to the total isospin T=0 defined by
\beq
A^+ = \sum_{i,j} \bar{x}_{ij} [P^+_i P^+_j]^{T=0} = \sum_{ij} x_{ij}
(P^+_{i,1} P^+_{j,-1}+P^+_{i,-1} P^+_{j,1}
           -P^+_{i,0} P^+_{j,0}).
\eeq
Supposing that the amplitudes $x_{ij}$ are separable, i.e., $x_{ij}=x_i x_j$, the collective 
quartet  operator can be written as
\beq
A^+= 2 \Gamma^+_1 \Gamma^+_{-1} - (\Gamma^+_0)^2,
\eeq
where $\Gamma^+_{t}= \sum_i x_i P^+_{i,t}$ denote, for t={0,1,-1},  the 
collective Cooper pair operators for the proton-neutron (pn), neutron-neutron (nn) and
proton-proton (pp) pairs. Thus, in this approximation
the state (2) can be written  as
\beq
| \Psi \rangle = (2 \Gamma^+_1 \Gamma^+_{-1} - \Gamma^{+2}_0 )^{n_q} |0 \rangle \nonumber 
         = \sum_k
\left(\begin{array}{l} n_q \\ k \end{array}\right)
(-1)^{n_q-k} 2^k (\Gamma^+_1 \Gamma^+_{-1})^k \Gamma^{+2(n_q-k)}_0 |0\rangle .
\label{psi}
\eeq
From the equation above it can be seen that the quartet condensate is a particular 
superposition of condensates of nn, pp and pn pairs. 

In Ref.\cite{qcm2} the quartet condensate model was  extended
to nuclei with  $N>Z$. 
For these nuclei it is   supposed that the neutrons in excess form a  pair 
condensate which is appended to  the quartet condensate. Thus, the ground state
of N $>$ Z nuclei is approximated by 
\begin{equation}
| \Psi \rangle = (\tilde{\Gamma}^+_1)^{n_N} (A^+)^{n_q} |0 \rangle = 
(\tilde{\Gamma}^+_1)^{n_N} (2 \Gamma^+_1 \Gamma^+_{-1} - \Gamma^{+2}_0 )^{n_q} |0 \rangle , 
\end{equation}
where $n_N=(N-Z)/2$ is the number of neutron pairs in excess and $n_q=(N-2n_N+Z)/4$ is 
the maximum number of alpha-like quartets which can be formed by the neutrons and protons.
Since the quartets $A^+$ have zero isospin, the state (6) has a well-defined total
isospin given by the excess neutrons, i.e., T=$n_N$.  
The neutron pairs in excess are described by the collective pair operator 
$\tilde{\Gamma}^+_{1}= \sum_i y_i P^+_{i1}$, which has a  different structure from 
the collective neutron pair entering in the collective quartet.
The mixing amplitudes $x_i$ and $y_i$ which define the ground state (6)
are determind from the minimization of $ \langle \Psi | H | \Psi \rangle$ under 
the normalization condition $ \langle \Psi | \Psi \rangle =1$. To calculate the
average of the Hamiltonian and the norm it is  used the recurrence 
relations method \cite{qcm1,qcm2}.

In Refs.\cite{qcm1,qcm2}  it was shown that  the quartet condensation model (QCM) 
presented above can describe with very  high accuracy (errors below 1$\%$)  the pairing correlations
 induced  by the isovector pairing force acting on a given single-particle spectrum.
This fact recommends QCM as a proper tool for treating the isovector pairing correlations
in self-consistent HF calculations. The calculation scheme we introduce here
is similar to the one commonly used in the HF+BCS calculations.  
Thus,  the isovector pairing force is employed
 as a residual interaction acting on the HF single-particle states from the  
vecinity of Fermi level.
In this study  the HF mean field is generated with   a zero range Skyrme functional   and the
HF  calculations are performed in a single-particle basis generated  by an axially deformed
 harmonic oscillator, as described in Ref.\cite{vautherin}.  After the HF calculations are converged,
 we select a set of neutron and proton single-particle states  with the energies located around the
 HF chemical  potentials. The energies of these states are considered in the Hamiltonian (1) for performing
the QCM calculations. Then, from the QCM calculations we extract the occupation probabilities of the 
pairing active orbits  which are further used to redefine the HF densities.  For example, the particle density for
neutrons and protons ($\tau=n,p$)  are defined by
\beq
\rho_{\tau}(r,z) = \sum_i v^2_{\tau, i} \| \psi_{\tau,i} (r,z) \|^2 ,
\eeq
where $v^2_{\tau,i}$ are the occupation probabilities for the single-particle 
states $\psi_{\tau,i}(r,z)$. They are taken equal to 1 (0) for the occupied 
(unoccupied) HF states which are not considered active in pairing calculations 
and equal to the QCM values otherwise. The HF and QCM calculations are iterated 
together until the convergence. Finally, the pairing energy is calculated by averaging 
the isovector pairing force on  the QCM state and is added to the mean-field energy.

As an illustration, the  HF+QCM calculation scheme outlined above is applied here for
studying the influence of isovector pairing correlations on symmetry and  Wigner energies.
In the phenomenological mass formulas these energies are parametrized by a quartic and, 
respectively, a linear term in $N-Z$. Thus, for an isobaric chain with A=N+Z  the ground 
state energy relative to the nucleus with  N=Z  can be written as
\beq
E(N,Z) = E(N=Z)+ a_A \frac{|N-Z|^2}{A} + a_W \frac{|N-Z|}{A}+  
\delta E_{shell} (N,Z) + \delta E_{P} (N,Z).
\eeq
In the equation above it is not considered the contribution of the Coulomb energy,
which is supposed to be  extracted from all the isotopes of the isobaric chain,
and it is also implicitely assumed that  for all nuclei with A=N+Z the volume and 
the surface energies are the same and therefore included in the term E(N=Z). 
The last two terms in Eq.(8) are the corrections associated to the shell structure
and pairing measured relative to the nucleus  with N=Z. Supposing that these two 
energy corrections can be also described by a linear and a quartic term in $|N-Z|$ and 
taking $T=|T_z|$, which is the case for the ground state of nuclei
with $A>40$, Eq.(8) can be written as 
\beq
E(T)=E(T=0)+\frac{T(T+X)}{2 \Phi},
\eeq
where we have used the notations of Ref.\cite{bentley1}.
In the equation above, associated sometimes with the concept of isorotational 
band \cite{frauendorf}, X quantifies the contribution of the linear
term in isospin to the ground state energy and takes into account all 
the possible effects, including the ones from the shell structure. 
The fit of Eq.(8) with experimental data shows that for many nuclei 
$a_A \approx a_W$. Thus, when the contribution of 
the last two terms of Eq.(8) are negligible, $X \approx 1$.
In this case the ground state energies of the isobaric chain relative
to the nucleus with N=Z depend on T(T+1), as the eigenvalues of the
total isospin $T^2$. However,  a systematic survey based on
experimental masses fitted with Eq.(9) \cite{bentley1} shows that 
$X$ is fluctuating quite strongly around $X=1$ (see Fig.3 below).
 
In what follows we analyze the prediction of the HF+QCM calculations
for the quantities $\Phi$ and $X$ of Eq.(9). The HF+QCM calculations are 
performed for isobaric chains of even-even nuclei with $A>40$ for which 
the ground state has $T=|T_z|$, as supposed in Eq.(9). 
For each isobaric chain the values of $\Phi$ and $X$ are extracted from the
binding energies of three nuclei with $T=|Tz|=0,2,4$, i.e., nuclei with 
N-Z=0,4,8. The Skyrme-HF calculations are done with the Skyrme functional
SLy4 \cite{sly4} and neglecting the contribution of the Coulomb interaction.  
The deformation is calculated self-consistently in  axial symmetry using
an harmonic oscillator basis \cite{vautherin}.
From the HF spectrum we  considered in the QCM calculations 10 single-particles
states, both for protons and neutrons, above a self-conjugate core. More precisely,
for an isobaric chain of mass A, with A/2 even, we chose a core with $N_c=Z_c=A/2-6$. 
For the N=Z nucleus this choice corresponds to the QCM state (5) with three quartets,
i.e., $n_q=3$. The same core is kept for calculating the other two isobars  with T=2,4. 
They are described by the QCM state (6) with $n_q=2, n_N=2$ and, respectively, $n_q=1, n_N=4$.

\begin{figure}
\begin{center}
\includegraphics*[scale=0.5,angle=-90]{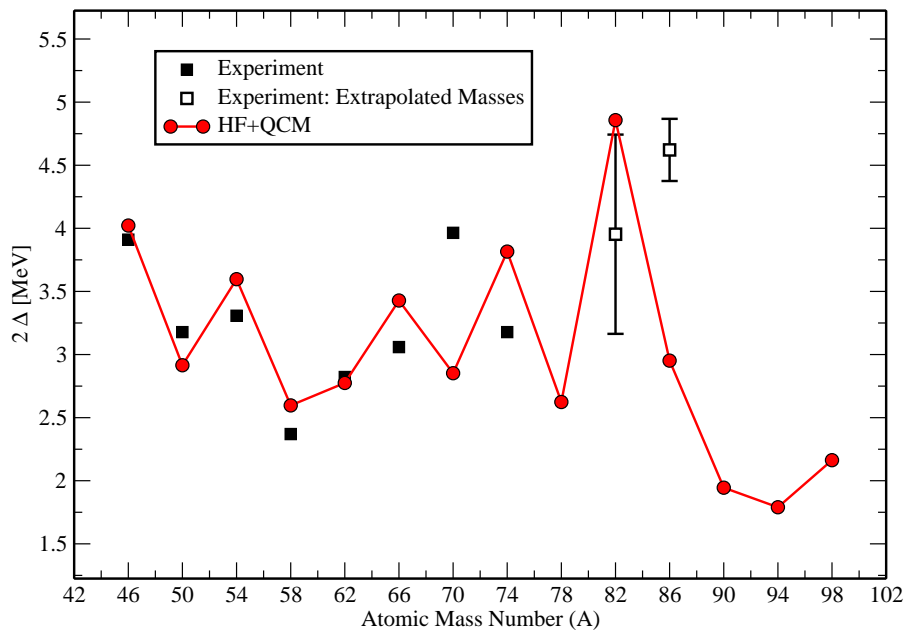}
\caption{ Even-even to odd-odd mass difference 
along the N=Z line calculated by Eq.(10) as a function of mass number. 
The experimental values, including the ones calculated from 
extrapolated masses, are from Ref.\cite{bentley1}. }
\end{center}
\end{figure}

In the HF+QCM calculations one needs a prescription for fixing the strength of
the isovector pairing force. According to Refs.\cite{vogel,macchiavelli}
the intensity of T=1 pairing can be measured by the 
difference in binding energies between the T=0 states in even-even
and odd-odd N=Z nuclei. This difference of binding energies is written as
\beq
2\Delta(N,Z)= \frac{B(N-1,Z-1)-2B(N,Z)+B(N+1,Z+1)}{2} ,
 \eeq
where N=Z=odd.
For odd-odd N=Z nuclei with $A>40$ the T=0 states one needs to employ 
in Eq.(10)  are  excited states.  The energies of these states are  
evaluated by adding to the neighboring even-even N=Z nuclei  a neutron 
and a proton in the single-particle orbits  just above the Fermi level and blocking them 
in the QCM calculations. In Fig.1 are shown the experimental $2\Delta$, extracted 
from Ref.\cite{bentley1}, in comparison with the HF+QCM results obtained with a state 
independent isovector force of strength $g=9.6/A^{3/4}$[MeV]. One can see that this pairing
force  gives a reasonable overall agreement with experimental data. Most likely
the largest deviations seen in Fig.1 are related to the crude approximation employed 
to calculate the excited T=0 states (e.g., the effect of the T=0 interaction is neglected) 
and to the inaccuracy of HF level densities around the Fermi levels (see below).

Figs.2-3 display the results of HF+QCM calculations for $1/\Phi$ and $X$ 
in comparison with the experimental values \cite{bentley1}.
The latter are obtained  employing in Eq.(9) the experimental masses 
of Ref.\cite{audi} from which the Coulomb energy was removed 
(for details, see \cite{bentley1}). In these figures are shown also the results
of HF+BCS calculations. In the BCS calculations, performed with the same model
space and cores as in the QCM calculations, the pairing correlations for protons 
and neutrons are treated independently and the proton-neutron pairing is not taken
into account.

\begin{figure}
\begin{center}
\includegraphics*[scale=0.5,angle=-90]{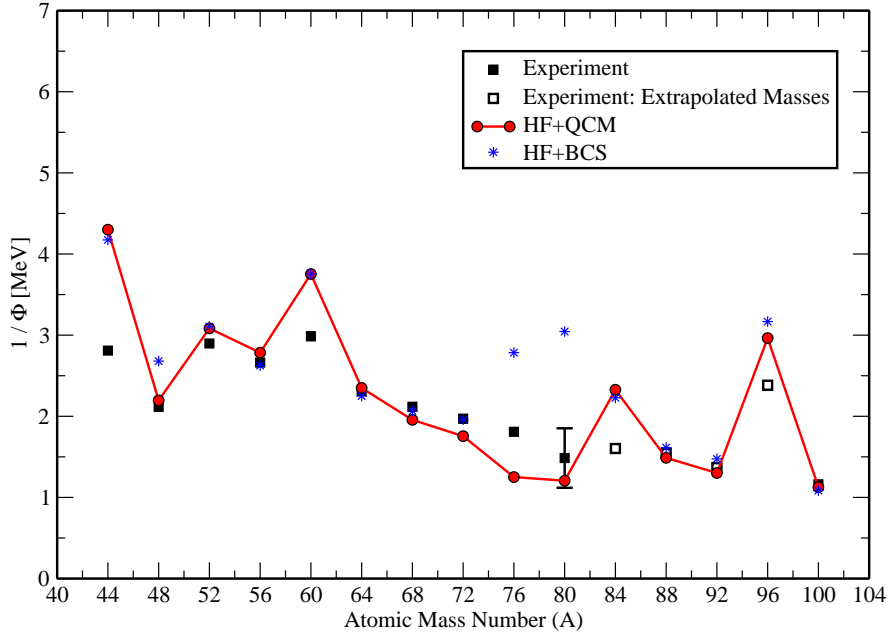}
\caption{ The quantity  $1/\Phi$ (see Eq.(9)), expressing the strength
of the symmetry energy term proportional to $T^2$, as a function of mass 
number. The experimental values, obtined by removing the contribution 
of Coulomb energy,   are from Ref.\cite{bentley1}.}
\end{center}
\end{figure}

From Fig.2 one can notice that the  HF+QCM
calculations describe  very well the  mass dependence of the quantity 
$1/\Phi$ associated to the standard symmetry energy proportional to $T^2$. 
The largest deviations appear for the isotopic chains which cross a magic 
number at T=2, i.e., for the nuclei with N-Z=4. The discrepancies are related 
to the inaccuracy of the deformations predicted by the mean field calculations 
for nuclei with two particles or two holes above/below a magic or semi-magic 
number. As an example we discuss here the result for the chain A=44 which
shows the largest deviation from the  experiment. The HF+QCM calculations 
predict almost zero deformation for $^{44}$Ti and $^{44}$Cr and a deformation of 
$\beta_2= 0.19$  for $^{44}$Ar. The exprimental deformations for these 
isotopes are, respectively, $\beta_2=0.268, 0.253, 0. 240$  \cite{raman}. 
Thus, contrary to what the experimental deformations indicate, in the calculations 
there is a large energy gap between the shells $f_{7/2}$ and $d_{3/2}$ which means
that the energy difference E(T=4)-E(T=2) is overestimated. On the other hand,
the energy difference E(T=2)-E(T=0) is underestimated since in going from 
T=0 to T=2 the pairs are interchanged within the degenerate $f_{7/2}$ shell.
Consequently, because $1/\Phi=E(T=4)-E(T=2)-[E(T=2)-E(T=0)]$,
the calculated value of $1/\Phi$ for A=44 is largely overestimated 
compared to the experiment. A similar  mechanism applies  for the chains 
$A=60,84,96$. From Fig.2 we can also observe that HF+BCS  and  HF+QCM give 
quite similar results for $1/\Phi$, suggesting that the isovector proton-neutron
pairing does not play a major role for the standard symmetry energy. 

\begin{figure}
\begin{center}
\includegraphics*[scale=0.5,angle=-90]{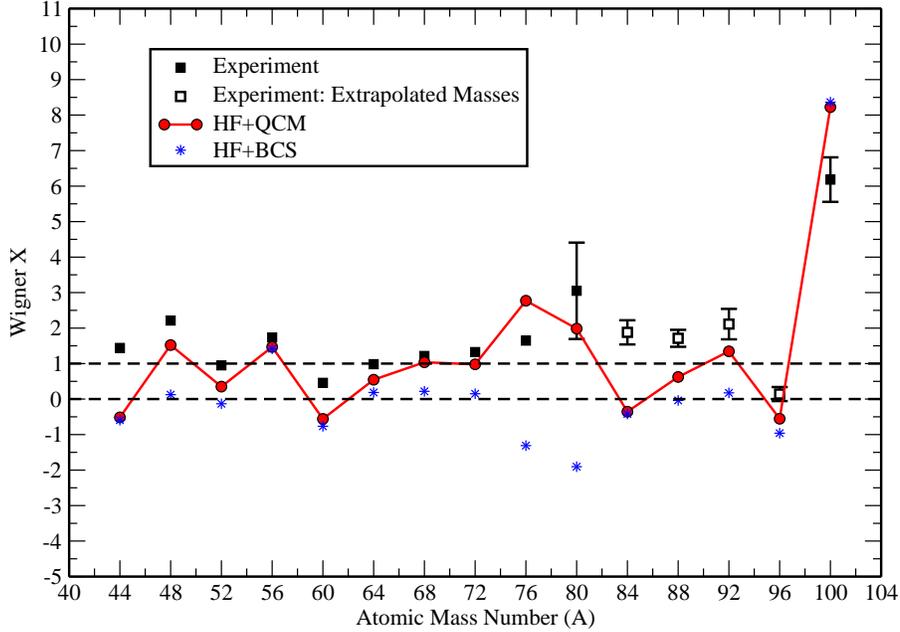}
\caption{ The quantity $X$ (see Eq.(9)), which gives the contribution
of Wigner energy relative to the standard symmetry energy, as a function 
of mass number. 
The experimental values, obtined by removing the contribution of 
Coulomb energy, are from Ref.\cite{bentley1}.}
\end{center}
\end{figure}

The predictions for the quantity $X$ are shown in Fig.3. 
One can now notice that  the HF+BCS calculations 
fail to describe the linear term in T associated to Wigner 
energy (see also Ref.\cite{satula}). In fact, as seen in Fig.3,
for the majority of chains the HF+BCS calculations predict
for X values close to zero. On the other hand we observe that 
the HF+QCM results are following   well the large 
fluctuations of $X$  with the mass number. The largest deviations 
from experimental values appear again for the isobaric chains 
which cross a magic number for T=2. It can be thus seen that for 
these chains the calculated 
$X$ values are underestimated (overestimated) when $1/\Phi$ are 
overestimated (underestimated). This fact can be simply traced back 
to the expression X=(3r-1)/(r-1), where r=(E(4)-E(2))/(E(2)-E(0)). 
For example, the underestimation of $X$ for the chain A=44 is due to 
the overestimation of the ratio $r$, which reflects the overestimation 
of $1/\Phi$ discussed above. Thus, as in the case of $1/\Phi$, the 
largest discrepancies of $X$ seen in Fig.3 are related to the 
inaccuracy of level densities predicted by mean field model for 
nuclei with two neutron or two holes above/below a magic number.

 A similar effect of the shell fluctuations on Wigner energy was noticed earlier 
 \cite{bentley1} by  using  a different calculation scheme  based on an isovector
 pairing force, digonalized in a restricted model space, and a phenomenological
 $T^2$ interaction introduced to simulate the isospin dependence of the 
 single-particle levels. It is  remarkably to 
 observe that the HF+QCM calculations,  in which the isospin dependence 
 of the  single-particle is consistently  taken  into account, give very good 
 results for the symmetry and Wigner energy terms,  comparable  in accuracy 
 with the  results  of Refs.\cite{bentley1,bentley2} obtained with additional 
 fitting parameters.
  
In conclusion, in this paper we have shown how the isovector pairing interaction 
can be treated in the mean-field models by conserving exactly the particle
number and the isospin. To treat the isovector pairing correlations we use a 
condensate of alpha-type quartets to which it is appended, in the case of nuclei
with $N>Z$, a condensate of neutron pairs. This formalism is applied to analyse the
effect of isovector pairing on symmetry and Wigner energies. The results show that
the isovector pairing acting on a self-consistent mean field can explain
reasonably well the mass dependence of Wigner energy. In principle, the latter
can be also influenced by the isoscalar proton-neutron pairing. This issue will 
be addresed in a future publication.

\vskip 0.4cm
\noindent
{\bf Acknowledgements}
\vskip 0.2cm
\noindent
N.S. thanks the hospitality of IPN Orsay, Universite Paris-Sud, where the article 
was mainly written. This work was supported by the Romanian Ministry of Education 
and Research through the grant Idei nr 57. D. N. acknowledges also the support 
from the PhD grant POSDRU/107/1.5/S/82514.

\end{document}